\documentclass{SCIS2019}

\newcommand{\ket}[1]{| #1 \rangle}

\begin{document}
\ArticleType{LETTER}
\Year{2019}
\Month{}
\Vol{}
\No{}
\DOI{}
\ArtNo{}
\ReceiveDate{}
\ReviseDate{}
\AcceptDate{}
\OnlineDate{}

\title{Experimental test of Tsirelson's bound with a single photonic qubit}{Experimental test of Tsirelson's bound with a single photonic qubit}

\author[1]{Zhiyu Tian}{}
\author[1,2]{Yuan-Yuan Zhao}{{zhaoyy63@mail.sysu.edu.cn}}
\author[1]{Hao Wu}{}
\author[1]{Zhao Wang}{}
\author[1]{Le Luo}{{luole5@mail.sysu.edu.cn}}

\AuthorMark{Z. Tian}

\AuthorCitation{Z. Tian, Y.-Y. Zhao, H. Wu, et al}


\address[1]{School of Physics and Astronomy, Sun Yat-Sen University, {\rm Zhuhai}, China}
\address[2]{Center for Quantum Computing,Peng Cheng Laboratory, {\rm Shenzhen518055}, China}

\maketitle

\begin{multicols}{2}
\deareditor
For many protocols, quantum strategies have advantages compared with their classical counter-partners, and these advantages have attracted many interests and applications. One of the famous examples is the Clauser-Horne-Shimony-Holt (CHSH) game~\cite{1}, which recasts Bell's theorem~\cite{2} into the framework of a game. In the CHSH game, two space-like separated players, Alice and Bob are each assigned a classical bit $a$ and $b$ respectively. Then they return bits $x$ and $y$ according to some pre-agreed strategies. They will win the game when $x\oplus y= a\cdot b$. In the game, if the players use the classical strategies, the optimal success probability $w(\text{CHSH})=0.75$. However, if they add some quantum resources, the success probability will increase and up to maximal value $cos^2(\pi/8)$, which is know as the Tsirelson's bound~\cite{3}. Moreover, Popescu and Rohrlich noted that the perfect success probability $1$ can also be achieved in a more general theory without violating the no-signaling assumption~\cite{4}.


Recently, a variant of the CHSH game, a so-called CHSH* game is proposed and theoretically investigated~\cite{5}. The CHSH* game is named for its similarity with the standard CHSH game. However, unlike the CHSH game involving two space-like separated parties and demanding conditions for its experimental realization, only a single player is involved in the CHSH* game. In this work, we study the two-dimensional CHSH* game with a single photonic qubit encoded in the polarization degree of freedom, and explore the success probability of the CHSH* game with different settings.


\lettersection{CHSH* game}
The CHSH* game involves only one player. As the Fig.~1(a) shown, the player applies the controlled transformation $A_a$ and $B_b$ on the input system in sequence and then performs measurement, obtaining an outcome $c$. The palyer wins if $c=a.b~(mod~2)$ and the optimal winning probability:
\begin{equation}
w(\text{CHSH*})=\max\limits_{all~strategies}\frac{1}{4}\sum_{a,b\in\mathbb{Z}_2}p(c=a.b|a,b).
\end{equation}

\lettersection{CHSH* game with a two dimensional quantum system}
For a $d=2$ quantum system in the unitary setting, without loss of generality, we set the initial state as $|+\rangle$ and the measurement as Pauli observable $X$. The optimal strategy can be obtained by optimizing the gates $A_a$ and $B_b$, where $a=0,1$ and $b=0,1$. As Ref.~\cite{5}, there exists one to one correspondence between the strategy for a CHSH game and a CHSH* game, and the Tsirelson's bound upper-bounds both of them by $A_0=\mathbb{I}$, $A_1=R_z(\frac{\pi}{2})$, $B_0=R_z(-\frac{\pi}{4})$ and $B_1=B_0^\dag$, where $R_z(\theta)$ represents the rotation around the $z$ axis. When varying $\theta$ in $B_b$ from $0$ to $\pi/2$, $P_{suc}$ changes with $\theta$ and $P_{suc}=1/2+sin\theta/4+cos\theta/4$. When $\theta=0,\pi/2$, $P_{suc}$ has the maximum probability for the Clifford setting, where $A_a$ and $B_b$ are confined to the Clifford gates.

At last, by including the irreversible transformation, such as the ERASE map which maps any qubit state to $\ket{0}$, in the setting of the CHSH* game, the player could win the game all the time~\cite{5}. A simple strategy attained $w(\text{CHSH*})=1$ is let the initial state be $|0\rangle$, the transformation $A_0=\mathbb{I}$, $A_1=X$, $B_0=ERASE$, $B_1=\mathbb{I}$, and the measurement be Pauli $Z$ observable.

\begin{figure*}[htbp]
    \centering
    \includegraphics[width=0.8\linewidth]{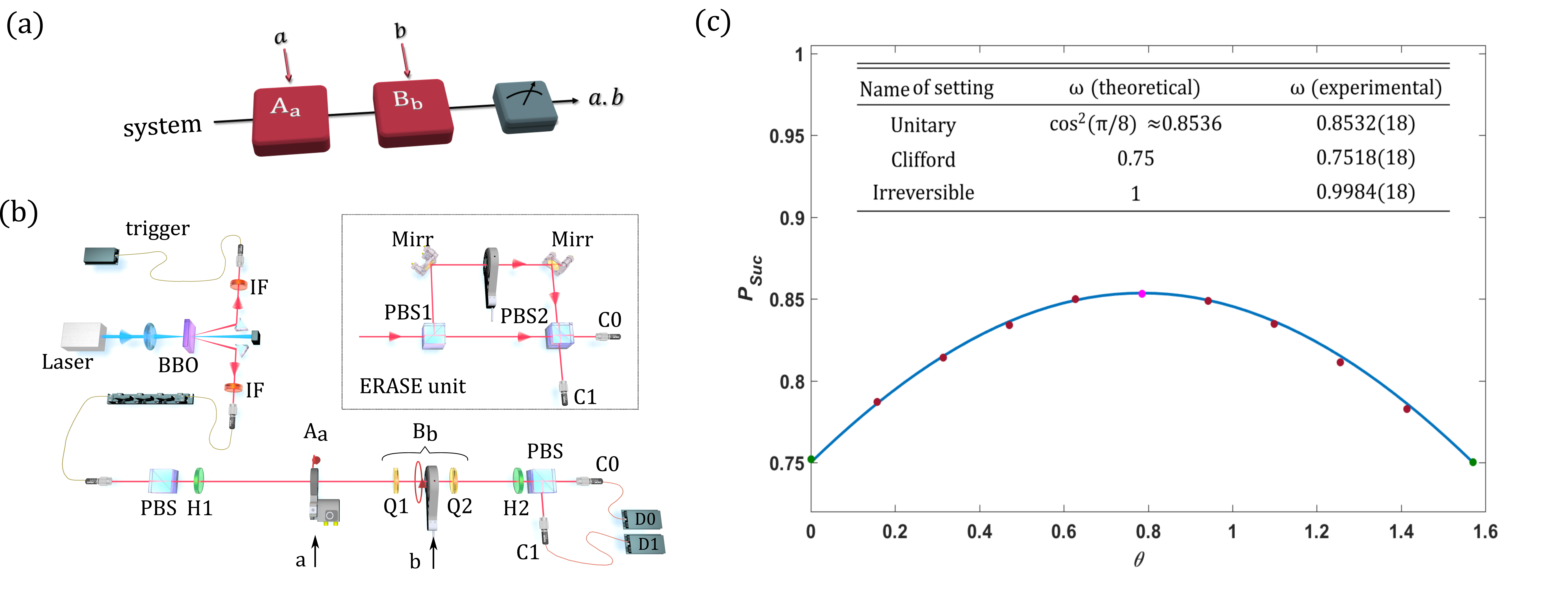}
    \caption{(a) The scheme of CHSH* game.
    (b) A $40~mW$, $404~nm$ wavelength continuous-wave laser pumps a type-I phase matching beta-barium-borate (BBO) crystal, and pairs of photons with $808~nm$ wavelength are generated in SPDC process. One of the photons is detected by the single photon detector as the trigger signal, and the other photon is sent to the CHSH* game modular through a single mode fiber. In the game, the photons are measured with the measurement $M$ after undergoing the controlled operations $A_a$ and $B_b$, then collected by the fiber couplers C0 and C1 and detected by the single-photon detectors D0 and D1. The measurement outcome $c$ is returned to check the winning condition. The details of the ERASE gate is given in the black box. Abbreviations: PBS, polarization beam splitter; Q, quarter wave plate; H, half wave plate; IF, interference filter; Mirr, mirror. (c) Result for the CHSH* game. The optimal success probabilities for the different settings are given in the Table. The dots represent the practical winning probabilities in our experiment and the blue line gives a theoretical prediction. The error bars which belong to 0.0017-0.0019, are smaller than the dot size.}
    \label{fig:1}
\end{figure*}

\lettersection{Experimental setup}
The CHSH* game is studied in a two-dimensional photonic system, which is spanned by two orthogonal polarized directions of the photon. As Fig.~\ref{fig:1}(b) shown, the single photon source is prepared by the spontaneous parametric down-conversion (SPDC) process. All the components in the setup are designed for the Gaussian beam light at $808~nm$ wavelength. To improve the precision of the operation, the spectrum of the down-converter photons is filtered using a $3~nm$ bandwidth interference filter (IF). A single mode fiber is used to direct the photon to the CHSH* game modular and ensure the zero-order Gaussian spacial mode.

In the game modular, a polarization beam splitter (PBS) combined with the $22.5^\circ$ rotated half wave plate (HWP) prepare the photon state into $\ket{+}=\ket{H}+\ket{V}$, where $\ket{H}$ and $\ket{V}$ represent the horizontally polarized direction and the vertically polarized direction respectively. We set the operation $A_0=\mathbb{I}$, $A_1=R_z(\pi/2)$, $B_0=R_z(-\theta)$, and $B_1=R_z(\theta)$, where $\theta\in[0,\pi/2]$. The switch between $A_1$ and $A_0$ is realized by flipping a quarter wave plate (QWP) into and out of the beam quickly. But for the operator $B_b$, a more general structure QWP-HWP-QWP must be applied. For this settings, the two QWPs' optical axis are oriented at $135^\circ$ and the translation between $B_0$ and $B_1$ is implemented by rotating the middle placed HWP between $\pi/4-\theta$ and $\pi/4+\theta$. At last, we perform the measurement $X$ on the single photon system with the $22.5^\circ$ HWP and the polarization beam splitter (PBS), and get $c=0$ when the detector $D0$ clicks, $c=1$ when the detector $D1$ clicks.

In our experiment, the QWP for $A_a$ and the middle HWP for $B_b$ are mounted in the motorized flip stage and the motorized rotation stage respectively, so the motion of them can be automated via the software interface. A Labview program is designed to monitor the whole process of the CHSH* game. For each run of the game, the control bit $a$ ($b$) is assigned a random number $0$ or $1$ in advance and the outcome $c$ is obtained by measuring the photon after undergoing the corresponding transformations $A_a$ and $B_b$. The winning probabilities $P_{suc}$ is shown in Fig.~\ref{fig:1}(c), where we test the case $\theta\in[0,\pi/2]$ with $\pi/20$ steps. For each dot, we run the experiment $1000$ times, and every time we collect about $38$ photons.

The irreversible setting is also studied in this 2D system. The initial state is prepared into $\ket{0}$ with a PBS and $A_0=\mathbb{I}$, $A_1=X$, $B_0=ERASE$, $B_1=\mathbb{I}$. An output-port conversion unit is needed to implement the switch between $B_0$ and $B_1$. In our experimental setup, the details of the unit can be found in the black box of Fig.~\ref{fig:1}(b). When the HWP mounted in the motorized rotation stage is rotated at $45^\circ$, it performs as a bit flip gate. The vertically polarized photon reflected at PBS1 becomes to $\ket{H}$ photon, transmits PBS2, and is then detected by the detector $D1$. When the original horizontally polarized photon directly transmits PBS1 and PBS2 and detected by the detector $D0$, the unit realizes an operation $\mathbb{I}$. In contrast, if the HWP is rotated at $0^\circ$, the photon will always be detected by $D0$ no matter which path it goes behind the PBS1, so the unit realizes an ERASE gate.


\lettersection{Results}
The CHSH* game is tested with a photonic qubit, and the results are given in Fig.~1(c). It shows that the maximal winning probabilities that can be obtained are sensitive to the settings of the game. When the game running with classical system and reversible gates, the optimal winning probability $w(\text{CHSH*})$ will only be $0.75$, which is same as the result of Clifford setting (green dots in Fig. 1(c)). However, the quantum source will serve an advantage. For the unitary gate, the winning probability increasing with $mod(\theta,pi/2)$, and the highest value in our experiment reaches $0.8536\pm0.0018$, approaching the Tsirelson's bound $w(\text{CHSH*})=w(\text{CHSH})=cos^{2}(\frac{\pi}{8})$. Furthermore, after removing the limit of irreversible transformations, we even win the game with an absolute probability $1$, which is $0.9984\pm0.0018$ for our experimental realization.

In conclusion, we investigate the CHSH* game with a single photonic qubit. For the reversible case, all the probabilities for $\theta\in(0,\pi)$ are higher than the classical upper bound $0.75$, moreover, the optimal quantum strategy($\theta=\pi/2$) can achieve the Tsirelson's bound. It is known that there exists a temporal version of the CHSH scenario, which probes the correlations of measurements happening at different times and can also yield the Tsirelson's bound with the single qubit system~\cite{6}. The quantum advantage of the temporal CHSH scenario comes from the violation of the assumption of `macroscopic realism' and `non-invasiveness'. However, in contrast to the spacial and temporal CHSH scenario, the CHSH* game doesn't involve any nonlocality property.  One work related to the quantum advantages in CHSH* game is Ref.~\cite{7}. In the paper, the authors introduce a new notion of contextuality for transformations, where the sequences of transformations are the contexts. According to the theorem in~\cite{7}, the average failure probability of the CHSH* game has an inequality relation with the noncontextuality. But the more specific answer about the quantum advantage in CHSH* game still needs further research.

In our work, we also detect the winning probability for the irreversible gate. These results help us to understand how the Tsirelson's bound arises in the strict physics condition and how the reversibility play a role in the advantages of quantum protocols. This work sheds light on the development of strategies in quantum information and computation.

In the future, it would be interesting to implement the CHSH* game with higher dimensions. As the proposition 6 in Ref.~\cite{5} points out, in the reversible setting with $d\geq3$, we can always achieve an optimal winning probability $w(\text{CHSH*})=1$. While $w(\text{CHSH*})=cos^{2}(\frac{\pi}{8})$ for the case of $d=2$, the CHSH* game can serve as a dimensional witness. The implementation will involve the generalized Pauli X, which can be realized with the method in Ref.~\cite{8,9} for photonic qudit. It is also possible to use the experimental setup of the CHSH* game to probe macrorealistic physical theories, such as testing Leggett-Garg inequalities in various quantum systems.

\lettersection{Acknowledgement}
We thank the fruitful discussions from Lorenzo Catani, Luciana Henaut, Shane Mansfield and Anna Pappa. The work is supported by the National Natural Science Foundation of China under Grant No.11804410, 11774436 and 11904423, Sun Yat-sen University Core Technology Development Fund, the Key-Area Research and Development Program of GuangDong Province under Grant No.2019B030330001, Guangdong Province Youth Talent Program under Grant No.2017GC010656, the Fund of Natural Science Foundation of Guangdong Province under Grant No.2017A030310452, Guangdong Basic and Applied Basic Research Foundation (Grants No. 2020A1515010864), Sun Yat-sen University Foundation for Youth (Grants No.17lgpy27), and National Training Programs of Innovation and Entrepreneurship for Undergraduates(Grants No.201901139).


\end{multicols}

\end{document}